\documentclass[12pt]{iopart}
\usepackage{bm}
\usepackage{subfigure}
\usepackage{inputenc}
\usepackage{iopams}
\usepackage{amsopn}
\usepackage{amssymb}
\usepackage{graphicx}
\usepackage{cite}
\usepackage{hyperref}
\DeclareMathOperator{\arctanh}{arctanh}

\begin{document}

%\fontfamily{phv}\selectfont

\title[Bandwidth effects in stimulated Brillouin scattering]{Bandwidth effects in stimulated Brillouin scattering driven by partially incoherent light} 
\author{B. Brand\~ao$^1$, J. E. Santos$^2$, R. M. G. M. Trines$^3$, R. Bingham$^{3,4}$, L. O. Silva$^1$}
%\email{luis.silva@tecnico.ulisboa.pt}
\address{$^1$GoLP/Instituto de Plasmas e Fus\~ao Nuclear, Instituto Superior T\'ecnico, Universidade de Lisboa, 1049-001 Lisboa, Portugal }
\address{$^2$Department of Applied Mathematics and Theoretical Physics, Centre for Mathematical Sciences, University of Cambridge, Wilberforce Road, Cambridge CB3 0WA, U.K.}
\address{$^3$Central Laser Facility, STFC Rutherford Appleton Laboratory, Didcot, OX11 0QX, U.K.}
\address{$^4$Department of Physics, SUPA, University of Strathclyde, Glasgow G4 0NG, U.K.}
% \newpage

\begin{abstract}
A generalized Wigner-Moyal statistical theory of radiation is used to obtain a general dispersion relation for Stimulated Brillouin Scattering (SBS) driven by a broadband radiation field with arbitrary statistics. The monochromatic limit is recovered from our general result, reproducing the classic monochromatic dispersion relation. The behavior of the growth rate of the instability as a simultaneous function of the bandwidth of the pump wave, the intensity of the incident field and the wave number of the scattered wave is further explored by numerically solving the dispersion relation. Our results show that the growth rate of SBS can be reduced by 1/3 for a bandwidth of 0.3 nm, for typical experimental parameters.
\end{abstract}

%\keywords{keyword1,keyword2,keyword3}

\submitto{\PPCF}
\maketitle

%%%%%%%%%%%%%%%%%%%%%%%%%%%%%%%%%%%%%%%%%%%%%%%%%%%%%%%%%%%%%%%%%%%%%%%

\section{\label{sec:1} Introduction}

\par All material substances interact nonlinearly with intense electromagnetic radiation leading to so-called parametric excitation or parametric instabilities \cite{armstrong62,Drake,Forslund,Kruer,background2}. In laser fusion parametric instabilities such as stimulated Brillouin and Raman scattering, filamentation and modulational instabilities \cite{armstrong62,Drake,Forslund} as well as self-focusing and plasma cavitation \cite{background3,background4} are detrimental to the coupling of the laser energy to the plasma. Stimulated Brillouin and Raman backscatter can result in a large fraction of the laser energy being scattered back out of the plasma before it reaches the critical surface, while filamentation of the laser beam creates beam break up resulting in hot spots and non-uniform illumination. To mitigate the effects of these parametric instabilities the use of broadband or incoherent lasers are being investigated. The standard treatment used to investigate these parametric instabilities use a coherent wave description of the laser which is limited when dealing with a broadband laser.

\par The use of the Wigner-Moyal statistical theory has proven to be powerful in studying these instabilities in nonlinear optics \cite{WP0.8}, demonstrating the stabilization of the modulational instability, as a result of an effect similar to Landau damping, driven by random phase fluctuations of the propagating wave. In similar studies \cite{WP0.9,WP0.10}, focusing on the onset of the transverse instability in nonlinear media in the presence of a partially incoherent light, the Wigner distribution was once more confirmed as a suitable approach. This formalism is particularly well suited for nonlinear optics because of the validity of the paraxial wave approximation, which justifies a forward propagating ansatz for the evolution of electromagnetic waves in dispersive nonlinear media. 

\par The Wigner-Moyal statistical approach to wave propagation has also enabled significant progress in the study of photon Landau damping \cite{photonlandau} and photon acceleration \cite{stochastic,titobook,resonant,trines2,trines3,yablon,wilks89}, where a time-dependent refractive index leads to a change in the frequency of electromagnetic waves (in contrast to a position-dependent refractive index, which leads to a change in wave number but not frequency). Both the modulational instability and photon acceleration have been extended to the study of drift waves interacting with zonal flows \cite{diamond1,diamond2,trines4,trines5,dodin2,dodin3,dodin4,dodin5}. More exotic applications include sea waves \cite{seawaves}, magneto-hydrodynamics \cite{weinberg}, dispersive Alfv\'en waves \cite{alfven1,alfven2} and neutrino-plasma interactions \cite{neutrino1,neutrino2,neutrino3}.

\par In laser-plasma interactions, in general, the standard Wigner-Moyal formalism is a limitation, as many critical aspects in ICF, fast ignition and several applications in laser-plasma and astrophysical scenarios demand a detailed analysis of the backscattered radiation. Early results on the scattering of electromagnetic waves by turbulent plasma were obtained by Bingham \textit{et al.} \cite{bingham_scatter}. In this paper we extend the work of Santos \textit{et al.} \cite{JorgePRL} where stimulated Raman scattering by a broadband pump was investigated using the Wigner-Moyal statistical approach to the investigation of stimulated Brillouin scattering.

\par The inclusion of bandwidth or incoherence effects in laser driven parametric instabilities has also been studied extensively using various approaches. The addition of small random deflections to the phase of a plane wave was shown to significantly suppress the three-wave decay instability \cite{WP0.7}, which was one of the first suggestions of the manipulation of the laser coherence as a way to avoid its deleterious effects. The threshold values for some electrostatic instabilities can also be effectively increased either by applying a random amplitude modulation to the laser or by the inclusion of a finite bandwidth of the pump wave \cite{WP0.1,WP0.5}. A new method for the inclusion of finite bandwidth effects on parametric instabilities, allowing arbitrary fluctuations of any group velocity, has also been developed \cite{WP0.4,WP0.3,pesme1,pesme2,pesme3,pesme4}. As far as Stimulated Raman Scattering is concerned, it became clear from these earlier works that, although it may seriously decollimate a coherent laser beam, laser bandwidth is an effective way to suppress the instability \cite{WP0.2}. The effects of laser beam incoherence induced by ``random phase plates'' have been studied extensively, and a reduction in the growth of many instabilities, including stimulated Raman and Brillouin scattering, has been demonstrated experimentally \cite{rpp1,rpp2,rpp3,rpp4,rpp5,rpp6}.

\par Mitigation of laser-plasma instabilities through increasing the bandwidth of the driving laser beam(s) has been investigated by several groups \cite{pesme3, follett1, hanwen,hansen1, zhao1, dorrer, bates}. For a parametric instability with ``coherent'' growth rate $\gamma_0$ and an incoherent pump laser with $\Delta\omega_0 \gtrsim \gamma_0$, Pesme \emph{et al.} \cite{pesme3} use an ``incoherent'' growth rate $\gamma_\mathrm{inc} = 4\gamma_0^2/\Delta \omega_0$. In theoretical studies \cite{pesme3,follett1,hanwen}, a bandwidth of $\Delta \omega_0 > 10\gamma_0$ or $\Delta \omega_0/\omega_0 \sim 5\%$ is often employed. In experimental studies \cite{hansen1, zhao1, dorrer, bates}, $\Delta \omega_0/\omega_0$ is typically much smaller, $\Delta \omega_0/\omega_0 < 1\%$, probably dictated by the properties of the intrinsic bandwidth of the laser gain medium.

\par Previous studies have also shown that a formalism that intrinsically describes the statistical properties of broadband lasers would allow for further theoretical progress and a systematic study on the control of parametric instabilities by spectral shaping of the pump laser. 

\par A statistical description of light can be achieved through the Wigner-Moyal formalism of quantum mechanics, which provides, in its original formulation, a one-mode description of systems ruled by Schr\"odinger-like equations. In order to address other processes apart from the direct forward scattering \cite{trines1}, a generalization of this Photon Kinetic theory (GPK) was developed in \cite{JorgeJMP}. This new formulation is completely equivalent to the full Klein-Gordon equation underpinning wave propagation in plasmas and was readily employed to derive a general dispersion relation for stimulated Raman scattering driven by white light \cite{JorgePRL}.

\par  In this paper, we focus on the study of the properties of stimulated Brillouin scattering (SBS) driven by a broadband pump. The suppression of the growth rate of the instability as a result of the inclusion of bandwidth in the pump wave is qualitative and quantitatively verified for realistic experimental parameters. For the sake of completeness, the less standard calculations are detailed in the appendices of the paper. 

\par This paper is organized as follows. In section \ref{sec:2} , we employ GPK to derive a general dispersion relation for SBS driven by a spatially stationary field with arbitrary statistics. We perform a detailed analytical study of different regimes of SBS and compare it with classical results for the monochromatic limit of the instability. For the first time, the whole domain of unstable wave numbers is numerically explored for a wide range of bandwidth choices. Finally, in section \ref{sec:4}, we summarize the main results and state the conclusions.   

\section{\label{sec:2} Broadband Stimulated Brillouin Scattering} 

\par We will start first by restating the fluid equation describing the plasma response, and the dependence of the driving term associated with the radiation on the plasma response from GPK, generalizing the equivalent result for monochromatic waves derived from the wave equation for the vector potential. These two equations will then be the basis to derive the dispersion relation relevant to the scenario under study.

\par In the following we use normalized units, where length is normalized to $c/\omega_{p0}$, with $c$ the velocity of light in vacuum and $\omega_{p0}=(4\pi e^2n_{e0}/m_ec^2)^{1/2}$ the electron plasma frequency, time to $1/\omega_{p0}$, mass and absolute charge to those of the electron, respectively, $m_e$ and $e$, with $e>0$. The plasma is modeled as an interpenetrating fluid of both electrons and ions, with $n_{e0}$ and $n_{i0}$ their equilibrium (zeroth order) particle densities, respectively. Densities are also normalized to the equilibrium electron density, such that the nornalized zero-order densities are $n_{e0}=1$ and  $n_{i0}=1/Z$, where $Z$ is the electric charge of the ions in units of $e$. 

\par Following the procedure outlined by Santos \textit{et al.} \cite{JorgePRL}, we define the normalized vector potential of the circularly polarized pump field as $\textbf a_p(\textbf r,t)=2^{-1/2}(\hat z+i\hat y)a_0\int d\textbf kA(\textbf k)$exp$[i(\textbf k.\textbf r-(\textbf k^2+1)^{1/2}t)]$, where $\textbf a_p=e\textbf A_p/m_ec^2$, $(\textbf k^2+1)^{1/2}\equiv\omega(\textbf k)$ is the monochromatic dispersion relation in a uniform plasma, and $\textbf A_p$ is the vector potential of the pump field.  We also allow for a stochastic component in the phase of the vector potential $A(\textbf k)=\hat{A}(\textbf k)$exp$[i\psi(\textbf r,t)]$ such that $\left<\textbf a_p^*(\textbf r+\textbf y/2,t).\textbf a_p(\textbf r-\textbf y/2)\right>=a_0^2m(\textbf y)$ is independent of $\textbf r$ with $m(0)=1$ and $|m(\textbf y)|$ is bounded between $0$ and $1$, which means that the field is spatially stationary i.e. the phase average of the pump field $\langle\ldots\rangle$ is not a function of $\textbf r$. In this section, $\tilde{q}$ denotes the first-order component of a generic quantity $q$. Unless specifically stated, the same notation for the functions and their Fourier transforms is used, as the argument of such functions (either $(\textbf r,t)$ or $(\textbf k,\omega)$) avoids any confusion. To obtain a dispersion relation for SBS we must couple the typical plasma response to an independently derived driving term, obtained within the GPK framework. 

\subsection{Plasma response and driving term}

\par In our previous work \cite{JorgePRL}, we studied the interaction of partially coherent light with electron plasma waves, whose (undriven, undamped) dispersion relation is given by $\omega_L^2 = 1 + (T_e/m_e) k_L^2$, with $T_e$ the electron temperature, thus covering stimulated Raman back- and forward scattering, and the relativistic modulational instability. In this work, we aim to study the interaction of partially coherent light with ion acoustic plasma waves, whose (undriven, undamped) dispersion relation is $\omega(k)^2 = (Z T_e/M) k^2$, with $M$ the ion mass. This allows us to study stimulated Brillouin back- and forward scattering, in both the weakly and strongly coupled regimes. We consider a fluid model for the plasma ion response to the ponderomotive force of the driving laser beam.

\par Combining the continuity and conservation of momentum equations for each species and closing the system with an isothermal equation of state for the electrons, we can readily present without more details the plasma response to the propagation of a light wave $\textbf a_p$, beating with its scattered component $\tilde{\textbf a}$, to produce the ponderomotive force of the laser, referring the reader to \cite{Drake,Forslund,Kruer,Silva99}: 

\begin{equation} \label{plasmaresponse}
\left(\frac{\partial^2}{\partial t^2}-2\tilde\nu\partial t-c_S^2\nabla^2\right)\tilde n=\frac{Z}{M}\nabla^2\mathrm{Re}[\textbf a_p.\tilde{\textbf a}],
\end{equation}
where $c_S\equiv\sqrt{Z T_e/M}$ is the ion sound velocity and $\tilde\nu$ an integral (damping) operator whose Fourier transform is $\nu|\textbf k_S|c_S$. Other models for the plasma response e.g. with more sophisticated descriptions of $\tilde\nu$ can be easily included in our analysis.   

\par We now need to describe how the incident pump wave and scattered radiation interact with the plasma. In standard formulations, the starting point is the wave equation for the vector potential with the corresponding source term given by the current associated with the plasma perturbations \cite{Drake,Forslund,Kruer,WP0.15}. 
In our approach, we derive the dependence of the driving term $\mathrm{Re}[\textbf a_p.\tilde{\textbf a}]$ on the plasma perturbation, using GPK. For the sake of completeness the derivation is given in \ref{AppendixDrivingTerm}. The driving term obtained within the framework of GPK is \cite{JorgeJMP,JorgePRL,Silva99}:

\begin{equation} \label{drivingterm}
W_{\mathrm{Re}\left[\textbf a_p.\tilde{\textbf a}\right]}=\frac 12\tilde n\left[\frac{\rho_0\left(\textbf k+\frac{\textbf k_S}{2}\right)}{D_s^-}+\frac{\rho_0\left(\textbf k-\frac{\textbf k_S}{2}\right)}{D_s^+}\right],
\end{equation}
where $W_{\mathrm{Re}\left[\textbf a_p.\tilde{\textbf a}\right]}$ represents the spatial and temporal Fourier transform of the Wigner function of $\mathrm{Re}\left[\textbf a_p.\tilde{\textbf a}\right]$. We observe that $W_f \equiv W_f (\textbf r,\textbf k , t)$, and thus the Fourier transform is $W_f (\textbf k_S,\textbf k , \omega_S)$, and $\omega_S$ and $\textbf k_S$ represent the frequency and wave vector of the ion acoustic wave. In Eq. (\ref{drivingterm}), $D_s^\pm$ is given by $D_s^\pm=\omega_S^2\mp\left[\textbf k.\textbf k_S-\omega_S\omega\left(\textbf k\mp\frac{\textbf k_S}{2}\right)\right]$, where we recall that $\omega$ is a function of $\textbf k$ via the linear dispersion relation $\omega(k)^2 = (Z T_e/M) k^2$.  

\subsection{General dispersion relation for Stimulated Brillouin Scattering and classical monochromatic limit}

\par We can now perform temporal and spatial Fourier transforms on the plasma response Eq. (\ref{plasmaresponse}), ($\partial t\rightarrow i\omega_S,\nabla_{\textbf r}\rightarrow -i\textbf k_S$), to obtain 
\begin{equation}
\tilde n=\frac ZM\frac{k_S^2}{\omega_S^2+2i\nu\omega_S|\textbf k_S|c_S-c_S^2\textbf k_S^2}\mathcal \mathrm{Re}[\textbf a_p.\tilde{\textbf a}],
\end{equation}
which can now be used with Eq.(\ref{drivingterm}). Taking advantage of one of the properties of the Wigner function \cite{WignerTransform.1,WignerTransform.2,WignerTransform.3,WignerTransform.4}
that states that
\begin{equation}
\int W_{f.g}d\textbf k=f^*g\Rightarrow\int\frac{W_{\mathrm{Re}\left[\textbf a_p.\tilde{\textbf a}\right]}}{\mathrm{Re}\left[\textbf a_p.\tilde{\textbf a}\right]}d\textbf k=1
\end{equation} 
we obtain the dispersion relation:
\begin{equation}
\fl 1=\frac{\omega_{pi}^2}{2}\frac{\textbf k_S^2}{\omega_S^2+2i\nu\omega_S|\textbf k_S|c_S-c_S^2\textbf k_S^2}\int \left[\frac{\rho_0\left(\textbf k+\frac{\textbf k_S}{2}\right)}{D_s^-}+\frac{\rho_0\left(\textbf k-\frac{\textbf k_S}{2}\right)}{D_s^+}\right]d\textbf k,
\end{equation}
%\end{footnotesize}
where $\omega_{pi}=\sqrt{Z/M}$ is the ion plasma frequency (in normalized units) and $f^*$ represents the complex conjugate of $f$. 

\par By making an appropriate change of variables, our general dispersion relation can be written in a more compact way as

\begin{equation}  \label{GeneralDispersionRelation}
1=\frac {\omega_{pi}^2}{2}\frac{\textbf k_S^2}{\omega_S^2+2i\nu\omega_S|\textbf k_S|c_S-c_S^2\textbf k_S^2} \int\rho_0(\textbf k)\left(\frac{1}{D^+}+\frac{1}{D^-}\right)d\textbf k ,
\end{equation} 
with $D^\pm(\textbf k)=[\omega(\textbf k)\pm\omega_S]^2-(\textbf k\pm\textbf k_S)^2-1$. Equation (\ref{GeneralDispersionRelation}) is the main result of this section. We observe that, given the statistical properties of the pump field, it is possible to evaluate Eq. (\ref{GeneralDispersionRelation}). This general dispersion relation can also be used to understand how spectral shaping can modify and mitigate Stimulated Brillouin Scattering. 

\par We first apply our general dispersion relation to the simple and common case of a pump plane wave of wave vector $\textbf k_0$, which means that $\rho_0(\textbf k)=a_0^2\delta(\textbf k-\textbf k_0)$. With the purpose of the following comparisons, we drop the contribution of the damping term $\nu=0$. The dispersion relation then becomes

\begin{equation}
\eqalign{
1=\frac {\omega_{pi}^2}{2}\frac{\textbf k_S^2}{\omega_S^2-c_S^2\textbf k_S^2}a_0^2\left\{\frac 1{[\omega(\textbf k_0)+\omega_S]^2-(\textbf k_0+\textbf k_S)^2-1}\right.+ \cr
%\end{equation}
%\begin{equation}
\phantom{1=}+\left.\frac 1{[\omega(\textbf k_0)-\omega_S]^2-(\textbf k_0-\textbf k_S)^2-1}\right\}.
}
\end{equation}

\par This result recovers the dispersion relation of Refs. \cite{Drake,Forslund,Kruer,lehmanndisp}, obtained for a coherent pump wave $\textbf A_S=\textbf A_{L0}\cos(\textbf k_0.\textbf r-\omega_0t)$, if we account for the difference in polarization and use $\omega_0=\omega(\textbf k_0)$. All the conclusions derived in Ref. \cite{Drake,Forslund,Kruer}, based on this dispersion relation, are then consistent with the predictions of GPK \cite{JorgeJMP}. 

\subsection{1D water-bag zero-order photon distribution function}

\par The full power of GPK becomes evident for broadband pump wave fields, where analytical results are not possible based on the standard formlism. In order to illustrate the consequences of broadband light, we consider a one-dimensional water-bag zero-order distribution function as the model for our photon distribution 

\begin{equation} \label{water-bag}
\rho_0(\textbf k)=\frac{a_0^2}{\sigma_1+\sigma_2}[\theta(k-k_0+\sigma_1)-\theta(k-k_0-\sigma_2)],
\end{equation}where $\theta(k)$ is the Heaviside function and $\sigma_1$ ($\sigma_2$) represents the spectral bandwidth to the left (right) of the central wave number, $k_0$.  
\par For this distribution function, the autocorrelation function of the random phase $\psi(x)$ satisfies

\begin{equation}
\left<\exp\left[-i\psi\left(x+\frac y2\right)+i\psi\left(x-\frac y2\right)\right]\right>=e^{-iy\tilde\sigma}\frac{\sin(y\bar\sigma)}{y\bar\sigma},
\end{equation}where $\tilde\sigma\equiv (\sigma_2-\sigma_1)/2$ and $\bar\sigma\equiv(\sigma_1+\sigma_2)/2$. The correlation length of this distribution is $\approx \pi/\sqrt 2\bar\sigma$.

\par A simplified dispersion relation for the water-bag distribution function of Eq. (\ref{water-bag}) can be derived (see \ref{AppendixWaterBag}) yielding

\begin{equation}
\eqalign{
1=\frac{a_0^2\omega_{pi}^2}{8\bar\sigma}\frac{k_S}{\omega_S^2-c_S^2k_S^2}\left[\frac{k_S^2}{k_S^2-\omega_S^2}\log\left(\frac{D_1^-D_2^+}{D_1^+D_2^-}\right)+\right. \cr
%\end{equation}
%\begin{equation} 
\phantom{1+}\left.+\frac{2\omega_S  k_S}{\sqrt{Q_0}}(\arctanh\textbf{ }b^++\arctanh\textbf{ }b^-)\right],
}
\label{water_bag_distribution_function}
\end{equation}
with $\omega_{0i}=\sqrt{[k_0+(-1)^i\sigma_i]^2+1}$, $D_i^\pm=\omega_S^2-k_S^2\pm 2[(k_0+(-1)^i\sigma_i)k_S-\omega_{0i}\omega_S]$, $Q_0=(k_S^2-\omega_S^2)(k_S^2-\omega_S^2+4)$, $Q^\pm=\prod_{i=1}^2[D_i^\pm+(k_S-\omega_S)(\omega_S\mp 2\omega_{0i})]$ and $b^\pm=2k_S^2(\omega_S+k_S)\sqrt{Q_0}(2\bar\sigma+\omega_{01}-\omega_{02})/\left[Q^0k_S^2-Q^\pm(\omega_S+k_S)^2\right]$.

\par We are interested in the maximum growth rate of SBS. Analytical results can be obtained in the case where all the photons of the distribution propagate in an underdense medium, which implies that $k_0+(-1)^i\sigma_i\gg 1$. This also guarantees that $k_0>\sigma_1$, which assures that $\rho_0(k)$ represents a broadband (pump) source of forward propagating photons, as expected. From this condition, the approximations $\omega_{0i}\approx k_0+(-1)^i\sigma_i$ and $b^\pm\approx 0$ are also valid.

\par The dispersion relation (\ref{water_bag_distribution_function}) then becomes
%\begin{small}
\begin{equation}
\eqalign{
1=\frac{a_0^2\omega_{pi}^2}{8\bar\sigma}\frac{k_S^3}{\omega_S^2-c_S^2k_S^2}\frac{1}{k_S^2-\omega_S^2}\left\{\ln\left[\frac{2(k_0-\sigma_1)+(\omega_S+k_S)}{2(k_0-\sigma_1)-(\omega_S+k_S)}\right]+\right. \cr
%\end{equation}
%\begin{equation}
\phantom{1=}\left.+\ln\left[\frac{2(k_0+\sigma_2)-(\omega_S+k_S)}{2(k_0+\sigma_2)+(\omega_S+k_S)}\right]\right\}.
}
\label{simplified}
\end{equation}
%\end{small}

\par In the weak coupling limit, $a_0^2 \ll 2 c_S k_S \omega_0 c_S^2/\omega_{pi}^2$, the dispersion of the plasma mode almost fully coincides with the ideal dispersion of an ion-acoustic plasma wave. 
The resonance condition for SBS can then be expressed as $\omega_S\sim k_Sc_S$, with $c_S\ll 1$ \cite{Drake,Forslund,Kruer}. Furthermore, the backscattering regime of stimulated Brillouin scattering (SBBS) is known to provide the highest growth rates \cite{Drake,Forslund,Kruer}, so we consider one of the terms $D_i^+$ resonant (corresponding to the contribution of the downshifted photons of the distribution function). By making the $D_1^+$ term resonant ($D_1^+=0\Rightarrow k_{L_{SBBS}}^m\approx 2(k_0-\sigma_1)/(1+c_S)$), we are considering the contribution of the photons of the lowest wave number, while with $D_2^+$ ($D_2^+=0\Rightarrow k_{L_{SBBS}}^M\approx 2(k_0+\sigma_2)/(1+c_S)$) we are searching for those of the highest wave number. This means that $k_S$ is of the order of $k_0$ and the range of unstable wave numbers is then given by:
\begin{equation} \label{range}
k_S\in\left[\frac{2}{1+c_S}(k_0-\sigma_1),\frac{2}{1+c_S}(k_0+\sigma_2)\right].
\end{equation}

\par We consider the upper limit case (as we will later see, the growth rate of the instability is within the same order of magnitude for the whole range of unstable wave numbers) and we note that $\omega_S\sim k_Sc_S$, with $c_S\ll 1$, implies that both $\omega_S\ll k_S$ and $\omega_S\ll k_0$. 

%\subsubsection{Weak field limit}

\par To determine the growth rate of the instability in the weak coupling limit, we now write $\omega \approx k_Sc_S+i\Gamma$, with $\Gamma$ being the real growth rate of the instability and $|\Gamma|\ll k_Sc_S$. The dispersion relation (\ref{simplified}) can then be rewritten in the form $1=A\ln B$ where

\begin{equation}
A=\frac{a_0^2\omega_{pi}^2}{8\bar\sigma}\frac{k_S^3}{\omega_S^2-c_S^2k_S^2}\frac{1}{k_S^2-\omega_S^2}\approx\frac{a_0^2\omega_{pi}^2(k_0+\sigma_2)}{4i(\sigma_1+\sigma_2)\Gamma c_Sk_S};
\end{equation}

\begin{equation}
\eqalign{
B=\frac{2(k_0-\sigma_1)+(\omega_S+k_S)}{2(k_0-\sigma_1)-(\omega_S+k_S)}\frac{2(k_0+\sigma_2)-(\omega_S+k_S)}{2(k_0+\sigma_2)+(\omega_S+k_S)}\approx \cr
%\end{equation}
%\begin{equation}
\phantom{B}\approx\frac{2k_0-\sigma_1+\sigma_2}{2(\sigma_1+\sigma_2)+i\Gamma}\frac{i\Gamma}{2(k_0+\sigma_2)}.
}
\end{equation}

\par We now take the imaginary part of the dispersion relation, working with a real $\Gamma$ and using the fact that, for a complex $Z=\rho e^{i\theta}$, with real $\rho$ and $\theta$, $\ln Z=\ln\rho+i\theta$. We get

\begin{equation} \label{general_case}
\Gamma c_Sk_S=\frac{a_0^2\omega_{pi}^2(k_0+\sigma_2)}{4(\sigma_1+\sigma_2)}\arctan\left[\frac{2(\sigma_1+\sigma_2)}{\Gamma}\right].
\end{equation}

\par With this result we are now able to compare our results for backscattering with those of Refs. \cite{Drake,Forslund,Kruer}. We found $k_{L_{SBBS}}^m\approx 2(k_0-\sigma_1)/(1+c_S)$ and $k_{L_{SBBS}}^M\approx 2(k_0+\sigma_2)/(1+c_S)$, which implies that, for the monochromatic limit, $k_{L_{SBBS}}^{m,pw}=k_{L_{SBBS}}^{M,pw}\equiv k_{L_{SBBS}}^{pw}= 2k_0/(1+c_S) \approx 2k_0(1-c_S)\approx 2k_0-2\omega_0c_S$, because $\omega_0\equiv\omega_{01}(\sigma_1=0)=\omega_{02}(\sigma_2=0)\approx k_0$, where we assume that the ion acoustic velocity is much smaller than the speed of light, $c_S \ll 1$. This recovers the result of Refs. \cite{Drake,Forslund,Kruer} for the wave number that maximizes the growth rate.

\par To determine the maximum growth rate in the weak coupling (\textit{wf}) scenario, we take the limit $\sigma_1,\sigma_2\rightarrow0$ and make use of $\arctan x\sim x$ when $x\rightarrow0$

\begin{equation} \label{weak_field_monochromatic}
%\Gamma_{SBBSwf}^{pw\text{,max}}=\frac{1}{2\sqrt 2}\frac{k_0\sqrt 2 a_0\omega_{pi}}{\sqrt{\omega_0k_0c_S}}.
\Gamma_{SBBSwf}^{pw,\mathrm{max}}=\frac{a_0\omega_{pi}}{2\sqrt{c_S}},
\end{equation}which also coincides with the monochromatic result in Refs. \cite{Drake,Forslund,Kruer} if we consider the already discussed correction for the polarization. 

\par We now go back to the general case of Eq. (\ref{general_case}) and work in the opposing limit, $(\sigma_1+\sigma_2)\gg \Gamma$, so the approximation $\arctan x\sim \pi/2- 1/x$ when $x\rightarrow\infty$ can be used, yielding

\begin{equation} \label{approximation_weak_field}
\Gamma_{SBBSwf}^{\mathrm{max}}=\frac{\pi a_0^2\omega_{pi}^2}{16c_Sk_0}\frac{k_0+\sigma_2}{\sigma_1+\sigma_2}
\left[1+\frac{a_0^2\omega_{pi}^2}{16c_Sk_0}\frac{k_0+\sigma_2}{(\sigma_1+\sigma_2)^2}\right]^{-1}.
\end{equation}

\par The corresponding saturation value for large bandwidth is

\begin{equation}
\Gamma_{SBBSwf}^{\mathrm{max},sat}=\frac{\pi a_0^2\omega_{pi}^2}{16c_Sk_0}.
\end{equation}

%\par We also recall the obtained limit for the monochromatic case (\ref{firstlimit}), now expressed in a simpler way,

%\begin{equation}\label{weak_field_monochromatic}
%\Gamma_{SBBSwf}^{pw,\text{max}}=\frac{a_0\omega_{pi}}{2\sqrt{c_S}}.
%\end{equation}

%\subsubsection{Strong field limit}

\par We now consider the strong coupling limit, i.e., we assume that $|\omega_S|\gg k_Sc_S$, which happens when $a_0^2 > 2 c_S k_S \omega_0 c_S^2/\omega_{pi}^2$ \cite{Drake,Forslund,Kruer}.
We work in the underdense limit, as in the weak coupling case, so that the range of unstable wave numbers still holds and we use $k_S\approx 2(k_0+\sigma_2)$ as the wave number for maximum growth, which means that $k_S$ is still of the order of $k_0$. We also neglect $|\omega_S|$ when compared to $k_0$, which establishes the scale $k_S c_S\ll |\omega_S|\ll k_S\approx k_0$, consistent with $c_S \ll 1$. This means that we are not neglecting the magnitude of the imaginary part of $\omega_S$ when compared to its real part.

\par We now expand $\omega_S=\alpha+i\beta$, with real $\alpha$ and $\beta$ and $|\alpha|, |\beta| \gg k_S c_S$, so that the dispersion relation yields (see  \ref{AppendixStrongFieldLimit})

\begin{equation}
\omega_S=\left(\frac{k_Sa_0^2\omega_{pi}^2}{2}\right)^{1/3}\left(\frac 12+\frac{\sqrt 3}{2}i\right),
\end{equation}
which is, once more, the result presented in Refs. \cite{Drake,Forslund,Kruer} with the usual polarization considerations. The maximum growth rate in the strong coupling limit is then

\begin{equation}
\Gamma_{SBBSsf}^{pw,\mathrm{max}}=\frac{\sqrt 3}{2}\left(\frac{k_Sa_0^2\omega_{pi}^2}{2}\right)^{1/3}.
\end{equation}

\subsection{Numerical solution of the complete dispersion relation}

\par We now examine the numerical solution of the complete dispersion relation in order to illustrate the evolution of the strength of the instability as a function of, not only the bandwidth, but also the wave number of the scattered wave itself.  

\par
\begin{figure}
\begin{center}\includegraphics[scale=0.55]{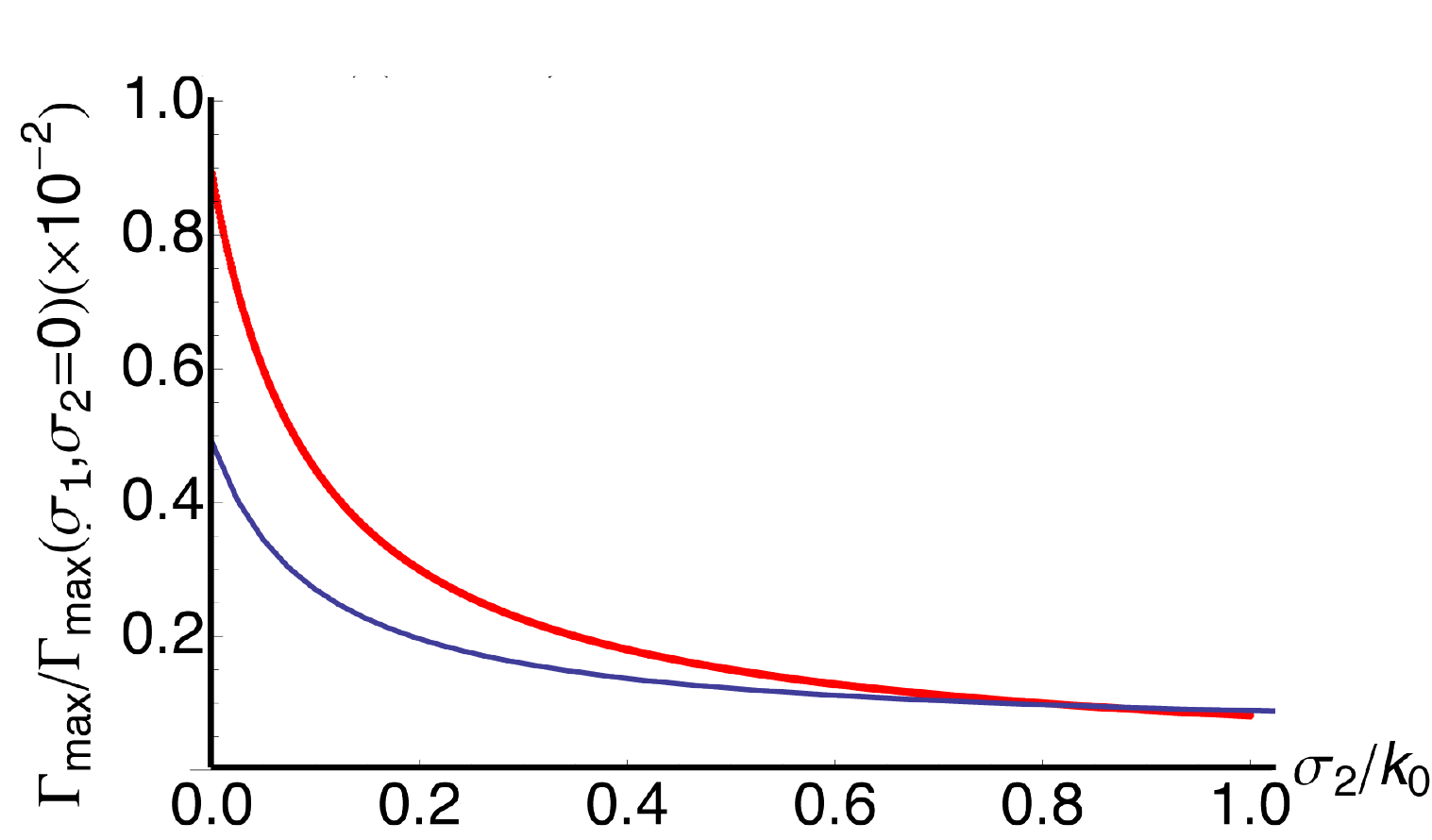}
\end{center}
\caption{Maximum growth rate of SBBS as a function of bandwidth - $a_0=0.1$, $k_0=80.0$, $\sigma_1=0.1k_0$, $c_S=0.01$, $\omega_{pi}=0.1$. Red line - numerical solution; blue line - analytical limit for $\Gamma\ll (\sigma_1+\sigma_2)$ of Eq. (\ref{approximation_weak_field})}\label{figuraB1}
\end{figure}

 In Fig. \ref{figuraB1} we show the maximum growth rate of the Brillouin instability as a function of the bandwidth parameter, $\sigma_2$, with $\sigma_1$ kept fixed. As expected, Eq. (\ref{approximation_weak_field}) is a good approximation to the complete solution only when we are dealing with large bandwidths. The difference between the approximate and the numerical solutions increases as bandwidth ($\sigma_2$) decreases.  As $\sigma_2$ approaches $k_0$, the results start to agree and Eq. (\ref{approximation_weak_field}) can be used. As we approach the monochromatic limit, only the numerical solution should be considered, as the choice of $\sigma_1=0.1k_0$ still accounts for a considerable difference between $\Gamma_{\mathrm{max}}(\sigma_2=0)$ and the maximum growth rate in the monochromatic limit,  $\Gamma_{\mathrm{max}}(\sigma_1,\sigma_2=0)$, expressed by Eq. (\ref{weak_field_monochromatic}). It is clear that a bandwidth as small as $10\%$ can still cause a reduction of the growth rate of the instability by a factor of more than $100$, which is significant.

\par
\begin{figure}
\begin{center}\includegraphics[scale=0.47]{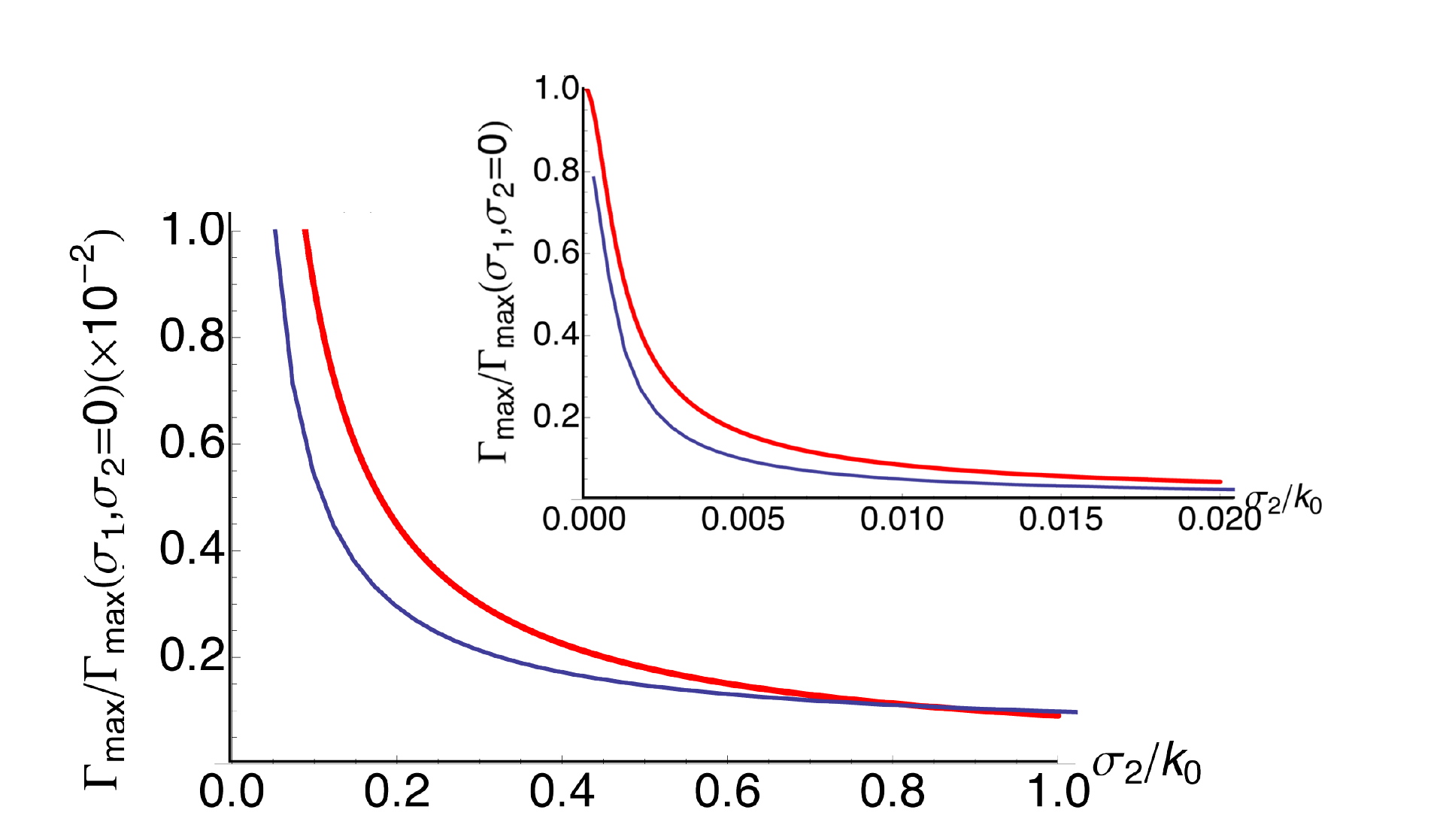}
\end{center}
\caption{Maximum growth rate of SBBS as a function of bandwidth - $a_0=0.1$, $k_0=80.0$, $\sigma_1\approx 0$, $c_S=0.01$, $\omega_{pi}=0.1$. Red line - numerical solution; blue line - analytical limit for $\Gamma\ll (\sigma_1+\sigma_2)$. In the inset the growth rate is shown for the regime where $\sigma_2/k_0\ll 1$}\label{figuraB2}
\end{figure}

 Fig. \ref{figuraB2} shows the same results for the case of $\sigma_2\approx 0$. As in the previous case, the approximation of Eq. (\ref{approximation_weak_field}) agrees with the numerical solution as $\sigma_2$ approaches $k_0$. The monochromatic limit of Eq. (\ref{weak_field_monochromatic}) can also be confirmed at the origin of the plot, as expected.  

\par
\begin{figure}
\begin{center}\includegraphics[scale=0.45]{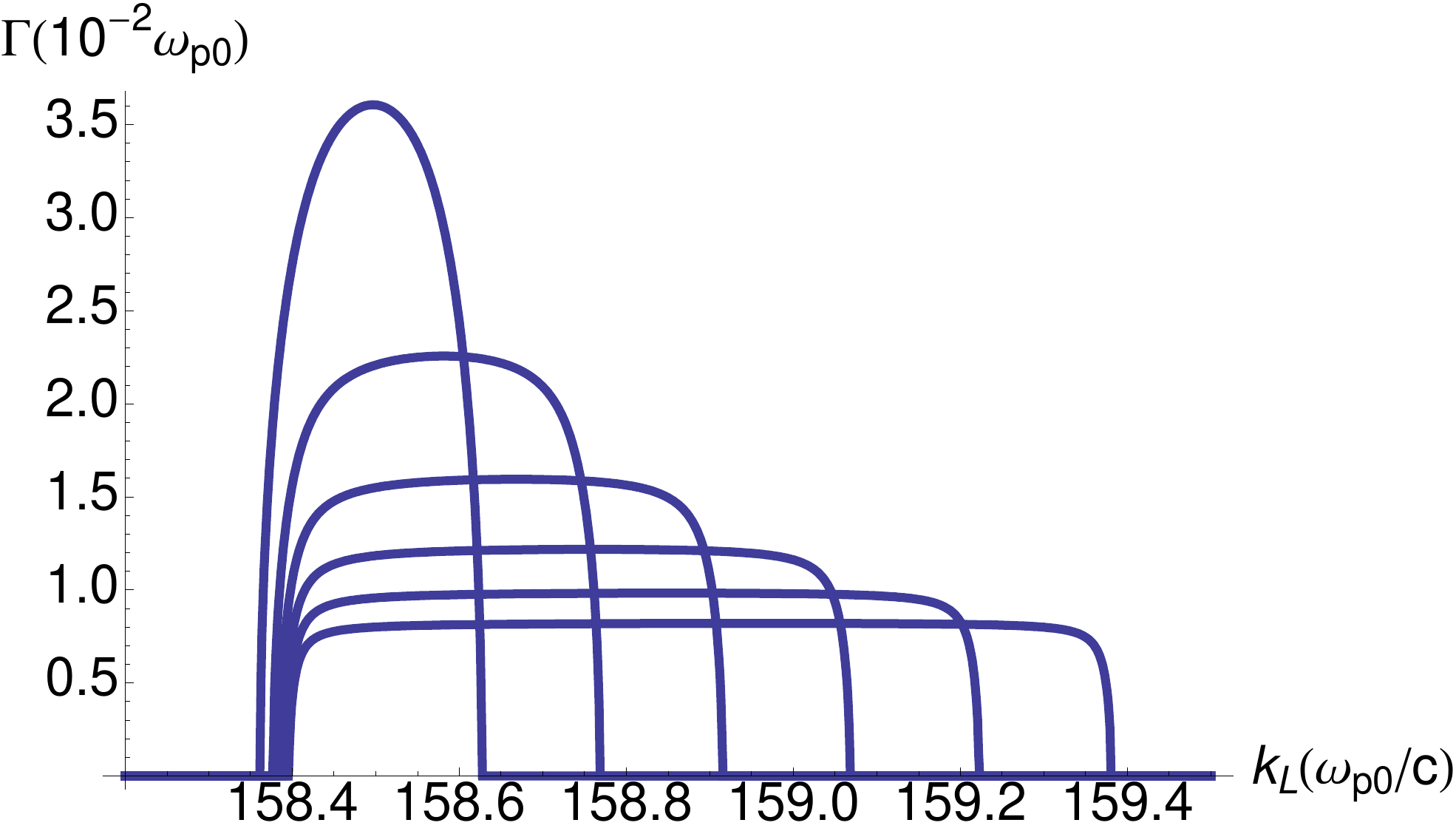}
\end{center}
\caption{Growth rate of SBBS as a function of the wave number of the scattered wave for different bandwidths of the water-bag (from the left to the right: $\sigma_2=0.1k_0,0.2k_0,0.3k_0,0.4k_0,0.5k_0,0.6k_0$, with $a_0=0.1$, $k_0=80.0$, $\sigma_1\approx 0$, $c_S=0.01$ and $\omega_{pi}=0.1$)}\label{figuraB6} 
\end{figure}

 We now study the behavior of the growth rate of the instability as a function of the wave number of the scattered wave. In Fig. \ref{figuraB6}, we plot the growth rate for a set of bandwidths and express it as a function of the wave number of the instability. We observe a very good agreement with the range of unstable wave numbers predicted by Eq. (\ref{range}). The lower limit does not depend on $\sigma_2$ and remains fixed as we increase bandwidth; as for the upper bound, it linearly grows as we increase the value of $\sigma_2$. 

\par We should also note that the flat structure observed indicates that the magnitude of the growth rate is within the same order for the full range of unstable wave numbers, meaning that the instability can grow on a wide range of wave numbers and lead to a significant level of ion acoustic turbulence. This is valid for relatively small bandwidths, as it is clear for $\sigma_2>0.1k_0$.  

\par
\begin{figure}
\begin{center}\includegraphics[scale=0.38]{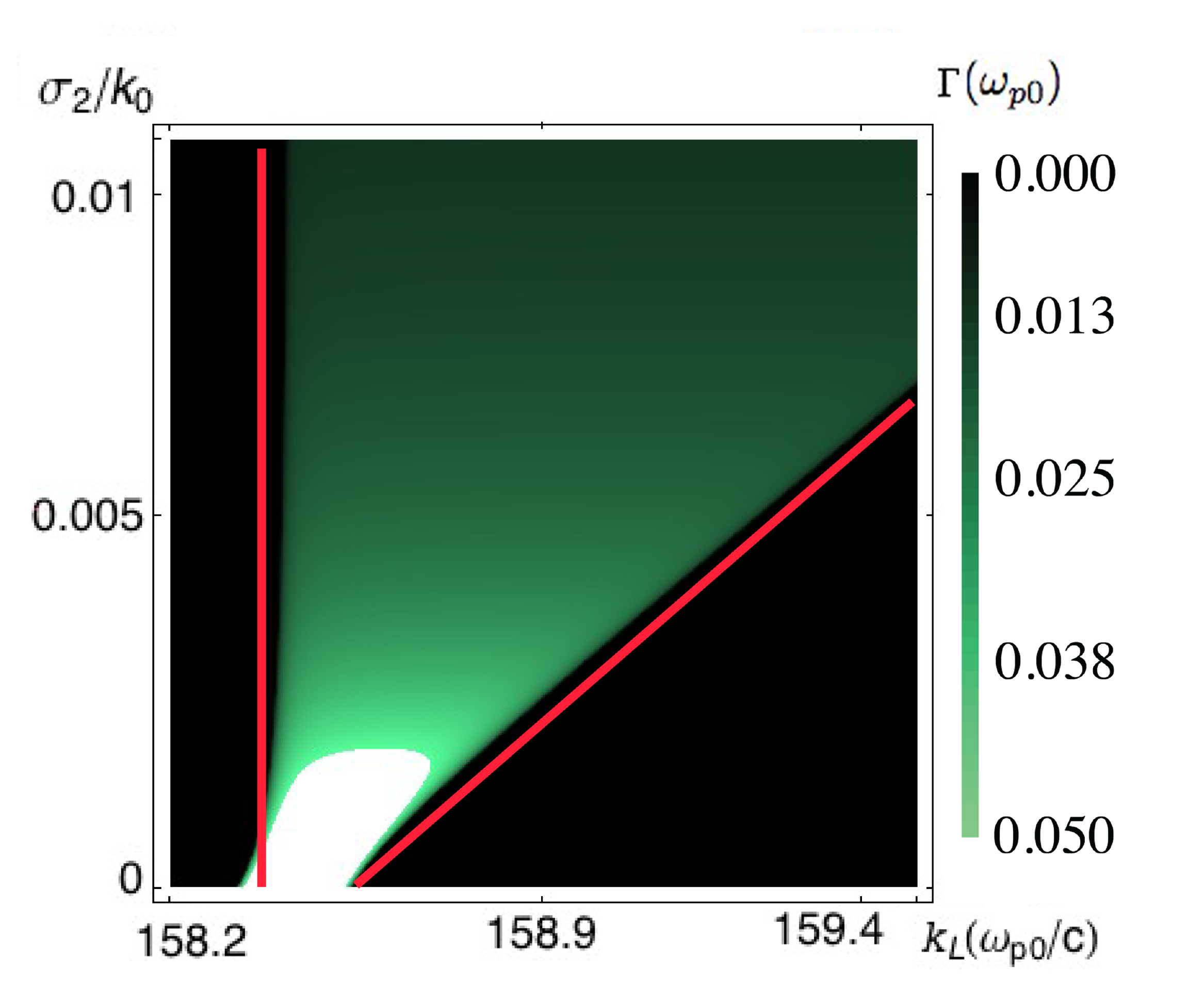}
\end{center}
\caption{Growth rate of SBBS as a function of the wave number of the scattered wave and the bandwidth of the water-bag: $a_0=0.1$, $k_0=80.0$, $\sigma_1\approx 0$, $c_S=0.01$, $\omega_{pi}=0.1$   (2D representation). The red lines illustrate the theoretical range of unstable wave numbers}\label{figuraB4}
\end{figure}

 In Fig. \ref{figuraB4}, the variation of the growth rate of SBBS as a continuous function of both the bandwidth of the pump and the instability wave number allows for a global picture of the instability. As expected, we observe a strong dependence of the instability on the bandwidth of the radiation used as a driver. For a bandwidth of just $1\%$ in $k_0$, the instability is already reduced to $10\%$ of the plane wave limit, which justifies the use of bandwidth as a means of significantly mitigating or reducing the growth of the instability. 

%\begin{figure}
%\begin{center}\includegraphics[scale=0.5]{Gamma_maxVsSigma_2_5}
%\end{center}
%\caption{Growth rate of SBBS as a function of the wave number of the scattered wave and bandwidth of the water-bag: $a_0=0.1$, $k_0=80.0$, $\sigma_1\approx 0$, $c_S=0.01$, $\omega_{pi}=0.1$  (3D representation)}\label{figuraB5} 
%\end{figure}

\par For fixed $k_0$, $a_0$ and $\sigma_1$, the growth rate for SBBS scales with $\propto 1/\sigma_2$, similarly to other distribution functions (e.g., asymmetric Lorentzian or Gaussian distribution of photons \cite{JorgePRL}). Both the wave number for maximum growth and the upper bound of the unstable wave numbers domain depend linearly on $\sigma_2$. 

%\par Even for small bandwidths, all the wave numbers in that range present a similar growth rate, suggesting the simultaneous excitation of the instability for an increasingly large set of wave numbers.
%\par For large values of $\sigma_2$, the growth rate for SBBS satures at the value $\frac{\pi a_0^2\omega_{pi}^2}{16c_Sk_0}$, which is then reached from above as the instability is suppressed.

\section{\label{sec:4} Conclusions}

\par A general dispersion relation for stimulated Brillouin scattering, driven by a partially coherent pump field, has been derived, using the GPK formalism \cite{JorgePRL} which is formally equivalent to the coupling of the full wave equation with the plasma fluid equations. After having retrieved the monochromatic limit of the equation, we have used a one-dimensional water-bag profile for the incident field to model broadband effects. The analysis has revealed a growth rate dependence on the coherence width $\sigma$ of the radiation field which scales with $1/\sigma$ typical of 3-wave processes \cite{JorgePRL}. Numerical estimates of the growth rate of the instability have been obtained as a function of the intensity of the incident field and the wave number of the scattered wave, confirming the theoretical predictions for the domain of unstable wave numbers.

\par The possibility of an accurate estimate of the growth rate of the instability, for a wide range of parameters, not only stresses the important role of bandwidth in the suppression of the instability, but also motivates an exploration of other photon distributions and a comparison with particle-in-cell simulations.

\par In this paper, we have focused on the backscattering regime of SBS, but the general dispersion relation we have derived (Eq. (\ref{GeneralDispersionRelation})) may be readily applied to different regimes. A detailed comparison with previous models for SBS pumped by a wave with finite bandwidth \cite{referenceChap3,referenceChap3_2,WP0.15,parabolicwaveequationapproximation1,parabolicwaveequationapproximation2} can then be performed and will be presented in the future, along with particle-in-cell simulations of parametric instabilities pumped by broadband radiation \cite{mythesis}.
A prediction of the suppression of SBS by the experimental mechanism of polarization smoothing \cite{polarizationsmoothing} or from other spectral distribution \cite{follett1} can also be be readily obtained through GPK and will be explored in future works. 

\par An oft-used method to increase the spectral bandwidth of a high-power laesr beam is spectral dispersion via random phase plates \cite{rpp1}. One of the issues with longitudinal and transverse smoothing by spectral dispersion on coherent laser beams is the creation of amplitude modulation resulting in enhanced intensity regions, seen as an adverse effect. A numerical study of beam smoothing by phase modulation on stimulated Brillouin scattering on the Laser M\'egajoule (LMJ) facility has been carried out \cite{duluc1,duluc2} and a pathway to a reduction of the amplitude modulation was demonstrated. Although it was shown that the effect can be reduced with the particular beam composition in LMJ, it is not obvious that this would be the case for all architectures and therefore the preferred route would be to develop broadband lasers from the outset.

\par Future laser drivers for inertial confinement fusion such as StarDriver \cite{star1} aim to control both laser plasma instabilities and hydrodynamic instabilities by using multiple beamlets with bandwidths in the range 2\%-10\%. Lasers with bandwidths of 2\% and repetition rates of 10Hz are already becoming available using Neodymium phosphate glass \cite{star1} and it is anticipated that this can be increased by using a range of laser gain media, such as a selection of Nd:Glass media based on phosphates, silicates and fluorides. Addition of other gain media such as Yb or Er glasses can potentially produce bandwidths up to 10\% \cite{star1}. In this paper and our previous paper, where the GPK formalism was used to study bandwidth effects on Raman scattering \cite{JorgePRL}, we demonstrated that lasers with bandwidths in this range significantly diminish the growth rate of both stimulated Brillouin and Raman backscatter.

%\par After having briefly described the contents and main fundamentals of this previous theory, we addressed its particular problem, while taking advantage of GPK's wide applicability. The appropriate translation of the geometry and most important features of the scenario described in \cite{referenceChap3} into our formalism allowed the simultaneous prediction of the growth rate of SBS by both models. Plots were made to compare these estimates for a set of qualitatively different scenarios and an excellent qualitative agreement was found. The monochromatic or classical limit was perfectly recovered by both models, and a suppression of the instability is always detected for increasing bandwidth of the pump wave. For significant values of bandwidth (while expressed in terms of the pump wave number), small divergences appear between the models and the previous agreement considerably deteriorates for larger spread values. 

\section*{Acknowledgements}
Work partially supported by the European Research Council (ERC-2015-AdG Grant no. 695088). RB acknowledges support from EPSRC grant EP/R004773/1. 
\appendix
%\section{Appendix} 

\section{Derivation of the driving term using Generalized Photon Kinetics}  \label{AppendixDrivingTerm}

\par The GPK derived in Ref. \cite{JorgePRL} can deal with the two mode problem, describing the radiation field \textbf a by two auxiliary fields $\mathbf \phi,\mathbf \chi=(\textbf a\pm i\partial_t\textbf a)/2$, thus allowing for a formally equivalent representation of the full wave equation in terms of two coupled Schr\"odinger equations for the auxiliary fields. With the introduction of four real phase-space densities

\begin{equation}
W_0=W_{\phi\phi}-W_{\chi\chi}
\end{equation}
\begin{equation}
W_1=2\mathrm{Re}[W_{\phi\chi}]
\end{equation}
\begin{equation}
W_2=2\mathrm{Im}[W_{\phi\chi}]
\end{equation}
\begin{equation}
W_3=W_{\phi\phi}+W_{\chi\chi}
\end{equation}
and the usual definition for the Wigner transform 
\begin{equation}
 W_{\textbf f.\textbf g}(\textbf k,\textbf r,t)=\left(\frac 1{2\pi}\right)^3\int \textbf f^*\left(\textbf r+\frac{\textbf y}{2}\right)\cdot \textbf g\left(\textbf r-\frac{\textbf y}{2}\right) \exp(i\textbf k \cdot \textbf y) d\textbf y   
\end{equation}
 as in Refs. \cite{WignerTransform.1,WignerTransform.2,WignerTransform.3,WignerTransform.4}, the coupled equations for $\mathbf\phi,\mathbf\chi$ (and, therefore, the complete Klein-Gordon equation) are shown \cite{JorgeJMP} to be equivalent to the following set of transport equations for the $W_i$, $i=0,...,3$

\begin{equation} \label{FirstTransportEquation}
\partial_tW_0+\hat{\mathcal{L}}(W_2+W_3)=0
\end{equation}
\begin{equation}
\partial_tW_1-\hat{\mathcal{G}}(W_2+W_3)-2W_2=0
\end{equation}
\begin{equation}
\partial_tW_2-\hat{\mathcal{L}}W_0+\hat{\mathcal{G}}W_1+2W_1=0
\end{equation}
\begin{equation} \label{LastTransportEquation}
\partial_tW_3+\hat{\mathcal{L}}W_0-\hat{\mathcal{G}}W_1=0
\end{equation}with the following definition for the operators $\hat{\mathcal{L}}$ and $\hat{\mathcal{G}}$

\begin{equation}
\hat{\mathcal{L}}\equiv\textbf k.\vec{\nabla}_{\textbf r}-n\sin\left(\frac 1 2\overleftarrow{\nabla}_{\textbf r}.\overrightarrow{\nabla}_{\textbf k}\right)
\end{equation}
\begin{equation}
\hat{\mathcal{G}}\equiv\left(\textbf k^2-\frac{\vec{\nabla}^2_{\textbf r}}{4}\right)+n\cos\left(\frac 1 2\overleftarrow{\nabla}_{\textbf r}.\overrightarrow{\nabla}_{\textbf k}\right)
\end{equation}where the arrows denote the direction of the operator and the trigonometric functions represent the equivalent series expansion of the operators. 
\par We first evaluate the zeroth order terms of each $W_i$, $i=0,...,3$, so we use $\textbf a =\textbf a_p$. It can be easily shown that

\begin{eqnarray}
W_{\phi\phi}^{(0)}&=&\frac{\rho_0(\textbf k)}{4}[1+\omega^2(\textbf k)+2\omega(\textbf k)]\\
W_{\chi\chi}^{(0)}&=&\frac{\rho_0(\textbf k)}{4}[1+\omega^2(\textbf k)-2\omega(\textbf k)]\\
W_{\phi\chi}^{(0)}&=&\frac{\rho_0(\textbf k)}{4}[1-\omega^2(\textbf k)]=-\frac{\rho_0(\textbf k)}{4}\textbf k^2
\end{eqnarray}where $\rho_0(\textbf k)\equiv W_{\textbf a_p.\textbf a_p}$ can be interpreted as the equilibrium distribution function of the photons.

\par We can immediately write

\begin{equation}
W_0^{(0)}=W_{\phi\phi}^{(0)}-W_{\chi\chi}^{(0)}=\rho_0(\textbf k)\omega(\textbf k)
\end{equation}
\begin{equation}
W_1^{(0)}=2\mathrm{Im}\left[W_{\phi\chi}^{(0)}\right]=0
\end{equation}
\begin{equation}
W_2^{(0)}=2\mathrm{Re}\left[W_{\phi\chi}^{(0)}\right]=-\frac{\rho_0(\textbf k)}{2}\textbf k^2
\end{equation}
\begin{equation}
W_3^{(0)}=W_{\phi\phi}^{(0)}+W_{\chi\chi}^{(0)}=\rho_0(\textbf k)\left(1+\frac{\textbf k^2}{2}\right)
\end{equation}where we have taken into account the reality conditions of the Wigner function \cite{WignerTransform.1,WignerTransform.2,WignerTransform.3,WignerTransform.4}.

\par We now explore the first order perturbative term of the transport equations  (\ref{FirstTransportEquation})-(\ref{LastTransportEquation}), as the zeroth order terms provide no new information: they are either trivial or equivalent to the dispersion relation for plane circularly polarized monochromatic waves in a uniform plasma, $\omega(\textbf k)=(\textbf k^2+1)^{1/2}$.  The first transport equation yields, in first order,

\begin{equation}
\partial_t\tilde W_0+\textbf k.\vec{\nabla}_{\textbf r}(\tilde W_2+\tilde W_3)-\tilde n\sin\left(\frac 1 2\overleftarrow{\nabla}_{\textbf r}.\overrightarrow{\nabla}_{\textbf k}\right)\rho_0(\textbf k)=0
\end{equation}

\par We now perform time and space Fourier transforms ($\partial t\rightarrow i\omega_S,\nabla_{\textbf r}\rightarrow -i\textbf k_S$), leading to

\begin{equation}
i\omega_S\tilde W_0-i\textbf k.\textbf k_S(\tilde W_2+\tilde W_3)+\tilde n\sin\left(\frac i 2 \textbf k_S.\vec\nabla_k\right)\rho_0(\textbf k)=0
\end{equation}

\par We note that we can write $\sin{\hat{\mathcal A}}=[\exp(i\hat{\mathcal A})-\exp(-i\hat{\mathcal A})]/(2i)$, for any operator $\hat{\mathcal A}$. Similarly, $\cos{\hat{\mathcal A}}=[\exp(i\hat{\mathcal A})+\exp(-i\hat{\mathcal A})]/2$. Making use of these relations, we have

\begin{equation}
e^{\textbf A.\mathbf\nabla_k}f(\textbf k)=\sum^\infty_{n=0}\frac{(\textbf A.\mathbf\nabla_k)^n}{n!}f(\textbf k)=f(\textbf k+\textbf A)
\end{equation}

\par The first transport equation can then be reduced to

\begin{equation}
\omega_S\tilde W_0-\textbf k.\textbf k_S(\tilde W_2+\tilde W_3)- \frac{\tilde n}{2} \left[\rho_0\left(\textbf k -\frac{\textbf k_S}{2}\right)-\rho_0\left(\textbf k +\frac{\textbf k_S}{2}\right)\right]=0
\end{equation}

\par We proceed analogously with the other three transport equations, leading to a system of four independent first order equations for the four variables $\tilde W_i$. 

\par We also note that
\[
%\begin{align}
W_2+W_3=W_{\phi\phi}+W_{\chi\chi}+2\mathrm{Re}[W_{\phi\chi}]=W_{\textbf a.\textbf a}
%\end{align}
\]

\par In zeroth order, as expected,

\begin{equation}
W_2^{(0)}+W_3^{(0)}=W_{\textbf a_p.\textbf a_p}=\rho_0(\textbf k)
\end{equation}

\par In first order, 

\begin{equation}
\tilde W_2+\tilde W_3=W_{\textbf a_p.\tilde{\textbf a}}+W_{\tilde{\textbf a}.\textbf a_p}=2W_{\textbf a_p.\tilde{\textbf a}}
\end{equation}where we have used the simmetry property of the Wigner distribution function that can be immediately derived from its realness ($W_{\textbf f.\textbf g}=W_{\textbf g.\textbf f}$).

\par We are only interested in a real electron density, so we take the real part of the right-hand side of the plasma response equation. Similarly, we write

\begin{equation} 
\tilde W_2+\tilde W_3=2W_{\mathrm{Re}\left[\textbf a_p.\tilde{\textbf a}\right]} \label{fifthequation}
\end{equation} 

\par We solve this equation together with the four independent equations for each $\tilde W_i$. The calculations are a bit lengthy but straightforward and yield

\begin{equation} 
W_{\mathrm{Re}\left[\textbf a_p.\tilde{\textbf a}\right]}=\frac{\tilde n}{2} \left[\frac{\rho_0\left(\textbf k+\frac{\textbf k_S}{2}\right)}{D_s^-}+\frac{\rho_0\left(\textbf k-\frac{\textbf k_S}{2}\right)}{D_s^+}\right],
\end{equation}

\begin{equation}
\frac{1}{D_s^\mp}=\frac{1\pm 2\textbf k.\textbf k_S/\omega_S^2 \pm (2\omega/\omega_S) \left(\textbf k+\frac{\textbf k_S}{2}\right)}{\omega_S^2-4\textbf k_S^2-\textbf k_S^2+4(\textbf k.\textbf k_S)^2/\omega_S^2-4}
\end{equation}

\par The expression for $D_s^\mp$ can be greatly simplified:
\[
%\begin{align}
D_s^\pm=\frac{(\omega_S^2\mp 2\textbf k\textbf k_S)^2-\left[2\omega_S\omega\left(\textbf k\mp\frac{\textbf k_S}{2}\right)\right]^2}{\omega_S^2\mp2\textbf k.\textbf k_S\mp 2\omega_S\omega\left(\textbf k+\frac{\textbf k_S}{2}\right)},
%\end{align}
\]
providing the final expression for the driving term,

\begin{equation} \label{drivingterm2}
W_{\mathrm{Re}\left[\textbf a_p.\tilde{\textbf a}\right]}=\frac 12\tilde n\left[\frac{\rho_0\left(\textbf k+\frac{\textbf k_S}{2}\right)}{D_s^-}+\frac{\rho_0\left(\textbf k-\frac{\textbf k_S}{2}\right)}{D_s^+}\right],
\end{equation}

\begin{equation}
D_s^\pm=\omega_S^2\mp\left[\textbf k.\textbf k_S-\omega_S\omega\left(\textbf k\mp\frac{\textbf k_S}{2}\right)\right].
\end{equation}

\section{Dispersion relation derivation for the one-dimensional water-bag distribution function}  \label{AppendixWaterBag}

\par Let $\rho_0(\textbf k)=a_0^2 [\theta(k-k_0+\sigma_1)-\theta(k-k_0-\sigma_2)]/(\sigma_1+\sigma_2)$, where $\theta(k)$ is the Heaviside function, in the generalized dispersion relation, so we get

\begin{equation} \label{justincase}
1=\frac{\omega_{pi}^2}{2}\frac{k_S^2}{\omega_S^2-c_S^2k_S^2}\frac{a_0^2}{\sigma_1+\sigma_2}\int_{k_0-\sigma_1}^{k_0+\sigma_2}\left[\frac{1}{D^+(k)}+\frac{1}{D^-(k)}\right]dk
\end{equation}with $D^\pm(k)=[\omega(k)\pm \omega_S]^2-(k\pm k_S)^2-1$, and $c_S=\sqrt{ZT_e/M}$. 

\par The integral of (\ref{justincase}) can be performed through the substitution $y=k-k_0$, so

\begin{equation}
\eqalign{
\int_{k_0-\sigma_1}^{k_0+\sigma_2}&\left[\frac{1}{D^+(k)}+\frac{1}{D^-(k)}\right]dk= \cr
%\end{equation}
%\begin{equation}
= \int_{-\sigma_1}^{\sigma_2} &\left[\frac{1}{2k_S y +2k_0k_S-(k_S^2-\omega_S^2) -2\omega_S\sqrt{(y+k_0)^2+1}}-\right. \cr
%\end{equation}
%\begin{equation}
&\left.-\frac{1}{2k_S y +2k_0k_S+(k_S^2-\omega_S^2) -2\omega_S\sqrt{(y+k_0)^2+1}}\right]dy
}
\end{equation}

\par We are then left with

\begin{equation}
I^\pm\equiv\int_{-\sigma_1}^{\sigma_2}\frac{1}{y+b^\pm+k\sqrt{(y+a)^2+1}} dy
\end{equation}
with $b^\pm\equiv k_0\pm(\omega_S^2-k_S^2)/(2k_S)$, $k\equiv-\omega_S/k_S$ and $a\equiv k_0$.

\par We start with the substitution $\sqrt{(y+a)^2+1}=y+t$, from which we obtain:
\begin{equation}
\eqalign{
y=\frac{a^2+1-t^2}{2(t-a)},\quad \frac{dy}{dt}=\frac{-4t(t-a)-2(a^2+1-t^2)}{4(t-a)^2}, \cr
\sqrt{(y+a)^2+1}=\frac{1+(t-a)^2}{2(t-a)}.
}
\end{equation}

\par The integral becomes

\begin{equation}
\mkern-36mu \int_{\sqrt{(a-\sigma_1)^2+1}+\sigma_1}^{\sqrt{(a+\sigma_2)^2+1}-\sigma_2}\frac{-4t(t-a)-2(a^2+1-t^2)}{2(t-a)\left[(a^2+1-t^2)+2(t-a)b^\pm+k(1+(t-a)^2)\right]}dt
\end{equation}

\par We perform one last transformation, $t-a=z$, which yields for the integral

\begin{equation}
-\int_{(\sigma_1-a)-\sqrt{(a-\sigma_1)^2+1}}^{-(\sigma_2+a)+\sqrt{(a+\sigma_2)^2+1}}\frac{1+z^2}{z[(k+1)+2(b^\pm-a)z+(k-1)z^2]}dz
\end{equation}

\par The problem has now been reduced to the computation of an integral of a rational function, for which a primitive can be explicitly obtained

%\begin{scriptsize}
\begin{equation}
\eqalign{
\fl I^\pm=-\left\{\frac{2(a-b^\pm)k}{(k^2-1)\sqrt{(k^2-1)-(a-b^\pm)^2}}\arctan\left[\frac{-(a-b^\pm)+(k-1)z}{\sqrt{(k^2-1)-(a-b^\pm)^2}}\right]+\right. \cr
%\end{equation}
%\begin{equation}
\left.+\frac{\ln z}{k+1}+\frac{\ln[(k+1)+2(b^\pm-a)z+(k-1)z^2]}{k^2-1}\right\}^{\sqrt{(a+\sigma_2)^2+1}-(a+\sigma_2)}_{\sqrt{(a-\sigma_1)^2+1}-(a-\sigma_1)}
}
\end{equation}

%\end{scriptsize}

\par If we define the quantity $Q^0\equiv (k_S^2-\omega_S^2)(k_S^2-\omega_S^2+4)$ and recall the property $\arctanh(x)=-i\arctan(ix)$, we can rewrite $I^\pm$ as

%\begin{scriptsize}
\begin{equation}
\eqalign{
\fl I^\pm=\left\{\mp\frac{2\omega_Sk_S}{\sqrt{Q^0}}\arctanh\left[\frac{\pm(k_S^2-\omega_S^2)+2(k_S+\omega_S)z}{\sqrt{Q^0}}\right]+\frac{k_S}{\omega_S-k_S}\ln z+\right.\cr
%\end{equation}
%\begin{equation}
\left.+\frac{k_S^2}{k_S^2-\omega_S^2}\ln[(k_S-\omega_S)\mp((k_S^2-\omega_S^2)z-(k_S+\omega_S)z^2]\right\}^{L_2}_{L_1},
}
\end{equation}
%\end{scriptsize}
where $L_1\equiv\sqrt{(k_0-\sigma_1)^2+1}-(k_0-\sigma_1)$ and $L_2\equiv \sqrt{(k_0+\sigma_2)^2+1}-(k_0+\sigma_2)$.

\par We now study the terms of the integral one by one. The second term ($I^\pm_2$) may be neglected as the contributions for the dispersion relation exactly cancel (the term does not depend on $b^\pm\Rightarrow I^+_2=I^-_2$). As for the third term, we write the argument of the logarithm, with $s=\pm 1$, as

%\begin{scriptsize}
\begin{equation}
\fl -(k_S+\omega_S)z^2-s(k_S^2-\omega_S^2)z+(k_S-\omega_S)=-(k_S+\omega_S)(z-z_{01})(z-z_{02}),
\end{equation}
%\end{scriptsize}
where $z_{01}$ and $z_{02}$ are the roots of the argument and $s=+1$ for $b^+$ (first contribution) and $s=-1$ for $b^-$ (second contribution). So we have

\begin{equation}
z_{01,2}=-\frac{s(k_S^2-\omega_S^2)\pm\sqrt{Q^0}}{2(k_S+\omega_S)}
\end{equation}

\par The third contribution to the dispersion relation is of the form $\frac{k_S^2}{k_S^2-\omega_S^2}\ln D$, where

\begin{equation}
\eqalign{
D &\equiv\frac{\left[Z_2+\frac{(k_S^2-\omega_S^2)+\sqrt{Q^0}}{2(k_S+\omega_S)}\right]\left[Z_2+\frac{(k_S^2-\omega_S^2)-\sqrt{Q^0}}{2(k_S+\omega_S)}\right]}{\left[Z_1+\frac{(k_S^2-\omega_S^2)+\sqrt{Q^0}}{2(k_S+\omega_S)}\right]\left[Z_1+\frac{(k_S^2-\omega_S^2)-\sqrt{Q^0}}{2(k_S+\omega_S)}\right]}\times\cr
%\end{equation}
%\begin{equation}
&\quad\times\frac{\left[Z_1-\frac{(k_S^2-\omega_S^2)-\sqrt{Q^0}}{2(k_S+\omega_S)}\right]\left[Z_1-\frac{(k_S^2-\omega_S^2)+\sqrt{Q^0}}{2(k_S+\omega_S)}\right]}{\left[Z_2-\frac{(k_S^2-\omega_S^2)-\sqrt{Q^0}}{2(k_S+\omega_S)}\right]\left[Z_2-\frac{(k_S^2-\omega_S^2)+\sqrt{Q^0}}{2(k_S+\omega_S)}\right]}
}
\end{equation}
where $Z_i\equiv\omega_{0i}-[k_0+(-1)\sigma_i]$.

\par We focus on each fraction individually and write them as in the following example

%\begin{small}
\begin{equation}
\frac{2[\omega_{02}-(k_0+\sigma_2)](k_S+\omega_S)+(k_S^2-\omega_S^2)+\sqrt{Q^0}}{2[\omega_{02}-(k_0+\sigma_2)](k_S+\omega_S)-(k_S^2-\omega_S^2)-\sqrt{Q^0}}\equiv\frac{A_1+B_1}{A_2+B_2},
\end{equation}
%\end{small}
where $A_{1,2}\equiv\mp(\omega_S^2-k_S^2)+2\omega_{02}\omega_S-2(k_0+\sigma_2)k_S$  and $B_{1,2}\equiv2\omega_{02}k_S-2(k_0+\sigma_2)\omega_S\pm\sqrt{Q^0}$. 

\par It can now be easily shown that $A_1A_2=B_1B_2\iff\frac{A_1+B_1}{A_2+B_2}=\frac{A_1}{B_2}$, so
 
\begin{equation}
\eqalign{
&\frac{2[\omega_{02}-(k_0+\sigma_2)](k_S+\omega_S)+(k_S^2-\omega_S^2)+\sqrt{Q^0}}{2[\omega_{02}-(k_0+\sigma_2)](k_S+\omega_S)-(k_S^2-\omega_S^2)-\sqrt{Q^0}}=\cr
%\end{equation}
%\begin{equation}
&=\frac{-(\omega_S^2-k_S^2)+2\omega_{02}\omega_S-2(k_0+\sigma_2)k_S}{2\omega_{02}k_S-2(k_0+\sigma_2)\omega_S-\sqrt{Q^0}}
}
\end{equation}

\par The second fraction may be written as $\frac{A_1+B_2}{A_2+B_1}=\frac{B_2}{A_2}$, so the product of the first two fractions becomes

\begin{equation}
\frac{A_1}{B_2}\frac{B_2}{A_2}\equiv -\frac{D_2^+}{D_2^-}
\end{equation}where $D_2^\pm=\omega_S^2-k_S^2\pm2[(k_0+\sigma_2)k_S-\omega_{02}\omega_S]$.

\par Proceeding similarly with the second group of two fractions, the total contribution to the dispersion relation is
\begin{equation}
I^+_2-I^-_2=\frac{k_S^2}{k_S^2-\omega_S^2}\log\left(\frac{D_1^-D_2^+}{D_1^+D_2^-}\right),
\end{equation}
\begin{equation}
D_i^\pm\equiv\omega_S^2-k_S^2\pm 2[(k_0+(-1)^i\sigma_i)k_S-\omega_{0i}\omega_S].
\end{equation}

\par Finally, the first contribution is the sum of two terms of the form

%\begin{scriptsize}
\begin{equation}
\eqalign{
\mp\frac{2\omega_Sk_S}{\sqrt{Q^0}} \left\{ \arctanh\left[\frac{\pm(k_S^2-\omega_S^2)+2(k_S+\omega_S)[\omega_{02}-(k_0+\sigma_2)]}{\sqrt{Q^0}}\right] \right. \cr
%\end{equation}
%\begin{equation}
\qquad \left.-\arctanh\left[\frac{\pm(k_S^2-\omega_S^2)+2(k_S+\omega_S)[\omega_{01}-(k_0-\sigma_1)]}{\sqrt{Q^0}}\right]\right\}
}
\end{equation}
%\end{scriptsize}

\par We make use of the property $\arctanh(x)-\arctanh(y)=\arctanh[(x-y)/(1-xy)]$ and write $\bar\sigma=(\sigma_1+\sigma_2)/2$, so the contribution becomes

\begin{equation}
I^+_3-I^-_3=\frac{2\omega_Sk_S}{\sqrt{Q_0}}(\arctanh\textbf{ }b^++\arctanh\textbf{ }b^-),
\end{equation}
\begin{equation}
b^\pm=\frac{2k_S^2(\omega_S+k_S)\sqrt{Q_0}(2\bar\sigma+\omega_{01}-\omega_{02})}{Q^0k_S^2-Q^\pm(\omega_S+k_S)^2},
\end{equation}
\begin{equation}
Q^\pm=\prod_{i=1}^2 [D_i^\pm+(k_S-\omega_S)(\omega_S\mp 2\omega_{0i})].
\end{equation}

\par Collecting all the terms, we get the final dispersion relation for the water-bag zero-order photon distribution

\begin{equation}
\eqalign{
1=\frac{a_0^2\omega_{pi}^2}{8\bar\sigma}\frac{k_S}{\omega_S^2-c_S^2k_S^2}\left[\frac{k_S^2}{k_S^2-\omega_S^2}\log\left(\frac{D_1^-D_2^+}{D_1^+D_2^-}\right)+\right.\cr
%\end{equation}
%\begin{equation} 
\phantom{1=}\left.+\frac{2\omega_Sk_S}{\sqrt{Q_0}}(\arctanh\textbf{ }b^++\arctanh\textbf{ }b^-)\right],
}
\end{equation}

with $\omega_{0i}=\sqrt{[k_0+(-1)^i\sigma_i]^2+1}$, $Q_0=(k_S^2-\omega_S^2)(k_S^2-\omega_S^2+4)$, and $D_i^\pm$, $Q^\pm$ and $b^\pm$ as given above.
%with $\omega_{0i}=\sqrt{[k_0+(-1)^i\sigma_i]^2+1}$, $D_i^\pm=\omega_S^2-k_S^2\pm 2[(k_0+(-1)^i\sigma_i)k_S-\omega_{0i}\omega_S]$, $Q_0=(k_S^2-\omega_S^2)(k_S^2-\omega_S^2+4)$, $Q^\pm=\prod_{i=1}^2[D_i^\pm+(k_S-\omega_S)(\omega_S\mp 2\omega_{0i})]$ and $b^\pm=2k_S^2(\omega_S+k_S)\sqrt{Q_0}(2\bar\sigma+\omega_{01}-\omega_{02})/\left[Q^0k_S^2-Q^\pm(\omega_S+k_S)^2\right]$.

\section{Derivation of the growth rate in the strong coupling limit}  \label{AppendixStrongFieldLimit} 

For small values of $a_0$, i.e. $a_0^2 \ll 2 c_S k_S \omega_0 c_S^2/\omega_{pi}^2$, the instability growth does not influence the magnitude of $\omega_S$ much, and we can write $\omega_S = c_S k_S + i\Gamma$ with $|\Gamma| \ll c_S k_S$. However, for $a_0^2 > 2 c_S k_S \omega_0 c_S^2/\omega_{pi}^2$, the instability growth strongly modifies the dispersion of the ion-acoustic wave, and the magnitude of $\omega_S$ \cite{Drake,Forslund,Kruer}. For this case, we write $\omega_S=\alpha+i\beta$ for real $\alpha$ and $\beta$ with both $|\alpha|, |\beta| \gg c_S k_S$.

We work in the underdense limit as in the weak coupling case, so that the range of unstable wave numbers still holds and we use $k_S\approx 2(k_0+\sigma_2)$ as the wave number for maximum growth, which means that $k_S$ is still of the order of $k_0$. We also neglect $|\omega_S|$ when compared to $k_0$, which establishes the scale $k_Sc_S\ll |\omega_S|\ll k_S\approx k_0$, consistent with $c_S\ll 1$. This means that we are not neglecting the magnitude of the imaginary part of $\omega_S$ when compared to its real part. 

\par Applying the expansion $\omega_S=\alpha+i\beta$ to the dispersion relation (\ref{water_bag_distribution_function}), we get

%\begin{scriptsize}
\begin{equation}
\eqalign{
1=\frac{a_0^2\omega_{pi}^2}{4(\sigma_1+\sigma_2)}\frac{k_S}{\alpha^2-\beta^2+i2\alpha\beta}\times\cr
\phantom{1=}\times\ln\left\{\frac{(2k_0-\sigma_1+\sigma_2)
\left[\alpha^2+\beta^2+2\alpha(\sigma_1+\sigma_2)+i2\beta(\sigma_1+\sigma_2)\right]}{2(k_0+\sigma_2)\left[\left(\alpha+2(\sigma_1+\sigma_2)\right)^2+\beta^2\right]}\right\}
}
\end{equation}
%\end{scriptsize}

\par We need both the real and imaginary parts of this equation, from which we obtain the following system of equations

%\begin{scriptsize}
\begin{equation}
2\alpha\beta=\frac{a_0^2\omega_{pi}^2}{4(\sigma_1+\sigma_2)}k_S\arctan\left[\frac{2\beta(\sigma_1+\sigma_2)}{\alpha^2+\beta^2+2\alpha(\sigma_1+\sigma_2)}\right]
\end{equation}
\begin{equation}
\eqalign{
\mkern-36mu\alpha^2-\beta^2=\frac{a_0^2\omega_{pi}^2}{4(\sigma_1+\sigma_2)}k_S\times\cr
\mkern-36mu\times\ln\left\{
\frac{(2k_0-\sigma_1+\sigma_2) \left[\left(\alpha^2+\beta^2+2\alpha(\sigma_1+\sigma_2)\right)^2+(2\beta(\sigma_1+\sigma_2))^2\right]^{1/2}}{2(k_0+\sigma_2)\left[(\alpha+2(\sigma_1+\sigma_2))^2+\beta^2\right]}\right\}
}
\end{equation}
%\end{scriptsize}

\par These equations can be numerically solved for $\alpha$ and $\beta$ to obtain he maximum growth rate $\Gamma=\mathrm{Im}(\omega_S)=\beta$. However, we focus on the plane wave limit and analytically derive the maximum growth rate of SBBS, for which we have a classical result \cite{Drake,Forslund,Kruer}. 

\par The equation for the imaginary part becomes

\begin{equation}
\alpha(\alpha^2+\beta^2)=\frac{a_0^2\omega_{pi}^2}{2}k_0
\end{equation}where we have used $\arctan x\sim x$ when $x\rightarrow 0$. 

\par The equation for the real part is more complicated and we work under the conditions $\sigma_1=0$ and $\sigma_2\rightarrow 0$. Neglecting terms of $\mathcal O(\sigma_2^2)$ in the arguments of the logarithms, the following approximation for the equation is valid

\begin{equation}
\beta^2-\alpha^2\approx\frac{a_0^2\omega_{pi}^2}{4}\frac{4k_0\alpha+\alpha^2+\beta^2}{\alpha^2+\beta^2}
\end{equation}where we have used the expansion $\ln (1+x)\sim x$ for $x\rightarrow 0$. 

\par Plugging the result for the imaginary part into this last equation, we obtain

\begin{equation}
\beta=\sqrt 3\alpha
\end{equation}

\par Using the equation for the imaginary part again, we get

\begin{equation}
\alpha=\frac 12\left(\frac{k_Sa_0^2\omega_{pi}^2}{2}\right)^{1/3}.
\end{equation}

\par $\omega_S$ can finally be written as

\begin{equation}
\omega_S=\left(\frac{k_Sa_0^2\omega_{pi}^2}{2}\right)^{1/3}\left(\frac 12+\frac{\sqrt 3}{2}i\right)
\end{equation}

\newpage

\end{document}